\documentclass[12pt]{article}

\usepackage{latexsym,amsmath,amssymb,theorem,epsfig} 

\DeclareFontFamily{OT1}{rsfs10}{} 
\DeclareFontShape{OT1}{rsfs10}{m}{n}{ <-> rsfs10 }{} 
\DeclareMathAlphabet{\mathscript}{OT1}{rsfs10}{m}{n} 

\numberwithin{equation}{section}

\newcommand{\cp}[1]{{\mathbb C}{\mathbb P}^{#1}}

\newcommand{\ns}{\normalsize}

\newcommand{\bC}{{\mathbf C}}
\newcommand{\bN}{{\mathbf N}}
\newcommand{\bV}{{\mathbf V}}
\newcommand{\bX}{{\mathbf X}}
\newcommand{\bY}{{\mathbf Y}}
\newcommand{\bZ}{{\mathbf Z}}

\def\cC{{\mathcal C}}
\def\cE{{\mathcal E}}

\def\cL{{\mathcal L}}

\def\cN{{\mathcal N}}
\def\cO{{\mathcal O}}

\def\cS{{\mathcal S}}

\theoremstyle{plain} 

{\theorembodyfont{\rmfamily} }

\begin{document}


\begin{titlepage}

\vspace{-5cm}

\title{
   \hfill{\ns UPR-871T, OUTP-99-03P, IASSNS-HEP- 00-03} \\[1em]
   {\LARGE Small Instanton Transitions  
     in Heterotic M--Theory} \\[1em] } 
\author{
   Burt A.~Ovrut$^1$, Tony Pantev$^2$
   and Jaemo Park$^3$\\[0.5em]
   {\ns $^1$Department of Physics, University of Pennsylvania} \\[-0.4em]
   {\ns Philadelphia, PA 19104--6396}\\
   {\ns $^2$Department of Mathematics, University of Pennsylvania} 
   \\[-0.4em]
   {\ns Philadelphia, PA 19104--6395}\\ 
   {\ns $^3$School of Natural Sciences, 
    Institute for Advanced Study} \\[-0.4em]
   {\ns Princeton, NJ 08540}\\}
\date{}

\maketitle

\begin{abstract}

We discuss non--perturbative phase transitions, within the context of heterotic
M--theory, which occur when all, or part, of the wrapped five--branes in the
five--dimensional bulk space come into direct contact with a
boundary brane. These transitions involve the transformation of the
five--brane into a ``small instanton'' on the Calabi--Yau space at the
boundary brane, followed by the ``smoothing out'' of the small instanton into
a holomorphic vector bundle. Small instanton phase transitions change the
number of families, the gauge group or both on the boundary brane,
depending upon whether a base component, fiber component or both components
of the five--brane class are involved in the transition. We compute the
conditions under which a small instanton phase transition can occur and 
present a number of explicit, phenomenologically relevant examples.
 
\end{abstract}

\thispagestyle{empty}

\end{titlepage}


\section{Introduction:}


In fundamental work, Ho\v rava and Witten \cite{HW1, HW2} 
showed that chiral fermions can be
obtained from M-theory by compactifying $D=11$, $N=1$ supergravity on an
$S^{1}/Z_{2}$ orbifold. The resultant theory consists of an
eleven--dimensional ``bulk'' space with two ten--dimensional boundary fixed
planes, one at one end of the bulk space and one at the other. Furthermore, in
order for the theory to be anomaly free, these authors showed that there must 
be a $D=10$, $N=1$ $E_{8}$ Yang--Mills supermultiplet on each of the boundary
planes. Witten \cite{W} further demonstrated that, 
when compactified on a
Calabi--Yau manifold, realistic low energy parameters,
such as the $D=4$ Newton's constant, will occur only if the the orbifold
radius is substantially larger than the radius of the Calabi--Yau threefold.

Based on this observation, the effective five--dimensional theory arising from
the compactification of $D=11$, $N=1$ supergravity on a Calabi--Yau threefold
was constructed \cite{losw1, losw2}. 
This five--dimensional regime of M--theory was shown to be
described by a specific type of gauged $D=5$, $N=1$ supergravity
coupled to hyper and vector supermultiplets in the bulk space and
to gauge and matter supermultiplets on the two four--dimensional boundary planes. 
The gauge symmetry introduces ``cosmological'' potentials for 
hyperscalars of a very specific type, both in
the bulk space and on the two boundary planes. These potentials have exactly
the right form so as to support BPS three--branes as solutions of the
equations of motion. It was shown in \cite{losw1} that the minimal 
static vacuum of this
theory consists of two BPS three--branes, one located 
at one boundary plane
and one at the other. However, as discussed in \cite{nse}, 
more complicated
``non--perturbative'' vacua are possible which, in addition to the 
pair of
boundary three--branes, allow for one or more five--branes in the bulk space.
These five--branes are wrapped on holomorphic curves in the background
Calabi--Yau threefold. When compactifying, it is necessary to 
specify the $N=1$ supersymmetry preserving $E_{8}$ 
gauge ``instanton'' localized on the Calabi--Yau 
space at each boundary three--brane. Such gauge configurations satisfy the
Hermitian Yang--Mills equations. These instanton vacua are not
arbitrary, being required to be, among other things, consistent with
anomaly cancellation. The physical effect of a non--trivial gauge instanton is 
to break $E_{8}$ to a smaller gauge group, as well as to introduce 
chiral ``matter'' superfields on the associated boundary three--brane. 
It was shown in \cite{Don, UhYau} that Hermitian Yang--Mills instantons can be described in
terms of smooth, semi--stable, holomorphic vector bundles. Using, and extending, 
mathematical techniques introduced in \cite{FMW, FMW2, D, BJPS}, 
it was demonstrated 
in a series of papers \cite{don1, don2, don3} that 
there are phenomenologically relevant, 
anomaly free vacuum solutions. These vacua have 
three--families of quarks and leptons, as well as realistic 
grand unified groups, such as $SO(10)$ and $SU(5)$, 
or the standard gauge group, $SU(3) \times SU(2) \times U(1)$, 
on one of the boundary
three--branes. This brane is  called the ``observable'' brane. 
For simplicity,
we usually take the vector bundle to be trivial on the other boundary
three--brane, which, thus, has unbroken gauge group $E_{8}$. This is called the
``hidden'' brane. Generically, we find that such vacua contain additional 
five--branes  ``living'' in the bulk space and wrapped on holomorphic curves
in the background Calabi--Yau threefold.

To conclude, we have shown in \cite{losw1, losw2, nse} 
that a fundamental ``brane world'' emerges from
M--theory compactified on a Calabi--Yau threefold and an $S^{1}/Z_{2}$
orbifold. Typically, the Calabi--Yau
radius is of the order inverse $10^{16} GeV$, whereas the orbifold radius can
be anywhere from an order of magnitude larger to inverse 
$10^{12} GeV$ \cite{scales}. For
length scales between these two radii, 
this world consists of a five--dimensional $N=1$ supersymmetric bulk
space bounded on two sides by BPS three--branes. One of these boundary 
three--branes, the ``observable'' brane, has  a realistic gauge group 
and matter content. The other boundary three--brane is the ``hidden'' brane
which, in this paper, we will choose to have unbroken $E_{8}$ gauge group. 
In addition, there generically are wrapped five--branes ``living''
in the bulk space.  We refer to this five--dimensional``brane world'' 
as heterotic M--theory.

A wrapped BPS five--brane in the bulk space has a modulus corresponding to
translation of the five--brane in the orbifold direction. An important
question to ask is: What happens to a wrapped bulk five--brane in heterotic
M--theory when it is
translated across the bulk space and comes into direct contact with one of the
boundary three--branes? This is the question that we address in this paper.
For specificity, and because it is physically more interesting, 
we will consider ``collisions'' of a bulk five--brane with
the ``observable'' boundary three--brane. All of our results, however, apply
equally well to collisions with the ``hidden'' brane. We
will show the following. Upon contact with the boundary three--brane, the
wrapped five--brane disappears and its data is ``absorbed'' into a singular
``bundle'', called a torsion free sheaf, localized on the Calabi--Yau
threefold associated with that boundary three--brane. This singular, torsion
free sheaf is referred to as a ``small instanton''\cite{Wi2}. This small instanton can
then be ``smoothed'' out by moving in its moduli space to a smooth holomorphic
vector bundle. The physical picture is that the bulk five--brane disappears,
thus altering the instanton vacuum on the boundary three--brane. The altered
gauge vacuum generically has different topological data than the instanton
prior to the five--brane ``collision'' with the boundary brane. In particular,
the third Chern class of the associated vector bundle can change, thus
changing the number of quark and lepton families on the observable wall. That
is, the vacuum can undergo a ``chirality--changing'' phase transition.
Furthermore, the structure group of the vector bundle can change, thus
altering the unbroken gauge group on the boundary brane. That is, the vacuum
can undergo a ``gauge--changing'' phase transition. Whether the phase
transition is chirality--changing, gauge--changing, or both depends on the
topological structure of the bulk five--brane being absorbed. At least for the
smooth holomorphic vector bundles discussed in this paper, we find that
chirality--changing small instanton phase transitions only occur for specific
initial topological data, and are otherwise obstructed. Gauge--changing
transitions can always occur. All of the work presented in this paper is
within the context of compactification on elliptically fibered Calabi--Yau 
threefolds. Related discussions involving ``monad'' Calabi--Yau spaces were
given in \cite{EvKa}.

Specifically, in this paper we do the following. In Section 2, we review some
properties of elliptically fibered Calabi--Yau threefolds that we will use in
our discussion. In Section 3, we present an outline of the spectral cover
construction of smooth, stable (and, hence, semi--stable) 
holomorphic $SU(n)$ vector bundles over elliptically
fibered Calabi--Yau threefolds, and give some explicit examples. Section 4 is
devoted to discussing the two fundamental topological conditions that must be
satisfied in any realistic particle physics vacuum. These conditions are
associated with the requirements of anomaly cancellation and 
three--families of quarks and leptons. Chirality--changing small instanton
phase transitions are explicitly constructed in Section 5. It is shown that
these transitions are associated with ``absorbing'' all, or part, of the
base component of the bulk space five--brane class. The mathematical
structure of the associated small instanton is presented and 
a detailed discussion of the
conditions under which it can be ``smoothed'' to a vector bundle is
given. We compute the Chern classes both before and after the small
instanton transition and give an explicit formula for calculating the change
in the number of families. We present several explicit examples, one
transition involving the complete base component and the second involving 
only a portion of the base component of the five--brane class. In Section 6,
we discuss gauge--changing small instanton transitions. It is shown that these
transitions are associated with ``absorbing'' all, or part, of the pure fiber
component of the bulk space five--brane class. We present the mathematical
structure of the associated small instanton and show that this can always be
``smoothed'' to a reducible, semi--stable vector bundle. We construct 
the Chern classes both
before and after the small instanton transition and show that the number of
families is unchanged. It is demonstrated   , however, that the structure group of the
vector bundle changes from $SU(n)$ to $SU(n)\times SU(m)$ for restricted
values of $m$. Hence, the unbroken gauge group on the observable brane, which
is the commutant of the structure group, also changes in a calculable way. We
present an explicit example of this type of transition. Finally, in Section 7,
we present our conclusions.


\section{Elliptically Fibered Calabi--Yau Threefolds:}


In this paper, we will consider Calabi--Yau threefolds, $X$, that are 
structured as elliptic curves fibered over a base surface,
$B$. Specifically,  there is a mapping
$\pi: X\to B$ such that $\pi^{-1}(b)$ is a torus, 
$E_b$, for each point $b \in B$. We further require that 
this torus fibered threefold has a zero section; that is, there 
exists an analytic map $\sigma : B\to X$ that assigns to 
every element $b$ of $B$, an element $\sigma(b) \in E_b$. The
point $\sigma(b)$ acts as the zero element for the group law 
and turns $E_b$ into an elliptic curve.

A simple representation of an elliptic curve is given in the 
projective space $\cp{2}$ by the Weierstrass equation
\begin{equation}
zy^2=4x^3-g_2xz^2-g_3z^3
\label{eq:1}
\end{equation}  
where $(x,y,z)$ are the homogeneous coordinates of $\cp{2}$ and $g_2$,
$g_3$ are constants. The origin of the elliptic curve is located at
$(x,y,z)=(0,1,0)$. Note that near the origin $z \cong 4x^3$ and,
hence, has a third order zero as $x \longrightarrow 0$. 
This same equation can represent the elliptic 
fibration, $X$, if the coefficients $g_2$, $g_3$ in the Weierstrass equation
are functions over the base surface, $B$. The correct way to express this 
globally is to replace the projective plane 
$\cp{2}$ by a $\cp{2}$-bundle $P \to B$ and then
require that $g_2$, $g_3$ be sections of appropriate line
bundles over the base. If we denote the conormal bundle to the 
zero section $\sigma(B)$ by $\cL$, then $P = {\mathbb P}({\mathcal
O}_{B}\oplus \cL^{2} \oplus \cL^{3})$, where ${\mathbb P}(M)$ stands
for the projectivization of a vector bundle $M$. There is a hyperplane line
bundle ${\mathcal O}_{P}(1)$ on $P$ which corresponds to the divisor 
${\mathbb P}(\cL^{2}\oplus \cL^{3}) \subset P$ and the
coordinates $x,y,z$ are sections of
$\cO_{P}(1)\otimes\cL^2, \cO_{P}(1)\otimes
\cL^3$ and $\cO_{P}(1)$ respectively. 
It then follows from \eqref{eq:1} that the coefficients $g_2$ and $g_3$ are 
sections of $\cL^4$ and $ \cL^6$.

It will be useful in this paper to define new coordinates\footnote{To see that 
such coordinates exist, note that the cubic behavior of $z$ as $x \to
0$ implies that the restriction of ${\mathcal O}_{P}(1)$ to $X$ is
precisely the line bundle $\cO_{X}(3\sigma)$. Therefore, we may take $\bZ$ to
be the unique (up to scalar) section of ${\mathcal O}(\sigma)$ and
normalize it so that $z = \bZ^{3}$.}, $\bX$,$\bY$,$\bZ$, 
on $X$ by $x=\bX\bZ$, $y=\bY$ and $z=\bZ^3$. 

It follows that $\bX,\bY,\bZ$ are now sections of line bundles
\begin{equation}
\bX \sim \cO(2\sigma)\otimes\cL^2 , \qquad
\bY \sim \cO(3\sigma)\otimes \cL^3 , \qquad
\bZ \sim \cO(\sigma)
\label{eq:3}
\end{equation}
respectively. The coefficients $g_2$ and $g_3$ remain 
sections of line bundles
\begin{equation}
g_2 \sim \cL^4, \qquad 
g_3 \sim \cL^6
\label{eq:4}
\end{equation}
The symbol ``$\sim$'' simply means ``section of''.

The requirement that elliptically fibered threefold, $X$, be a
Calabi--Yau space 
constrains the first Chern class of the tangent bundle, $TX$, to
vanish. That is, 
\begin{equation}
c_1(TX)=0
\label{eq:5}
\end{equation}
It follows from this that 
\begin{equation}
\cL=K_B^{-1}
\label{eq:6}
\end{equation}
where $K_B$ is the canonical bundle on the base, $B$. 
Condition~\eqref{eq:6} is
rather strong and restricts the allowed base spaces of an elliptically
fibered Calabi--Yau threefold to be del Pezzo, Hirzebruch and Enriques
surfaces, as well as certain blow--ups of Hirzebruch 
surfaces \cite{Grassi, MoVa} .


\section{Spectral Cover Description of $SU(n)$ Vector Bundles:}


As discussed in detail in \cite{FMW, don2},  $SU(n)$ vector 
bundles over an elliptically
fibered Calabi--Yau threefold can be explicitly constructed from two
mathematical objects, a divisor $\cC$ of $X$, called the spectral cover, and a
line bundle $\cN$ on $\cC$. Let us discuss the relevant properties of each in
turn. In this section, we will describe only stable $SU(n)$ vector bundles
constructed from irreducible spectral covers. Semi--stable vector bundles
associated with reducible spectral covers will be discussed in Section 6.

\subsection*{Spectral Cover:}

A spectral cover, $\cC$, is a surface in $X$ that is an $n$-fold cover of the
base $B$. That is, $\pi_{\cC}: \cC\to B$. The general form for a
spectral cover is given by
\begin{equation}
\cC=n\sigma + \pi^*\eta
\label{eq:7}
\end{equation}
where $\sigma$ is the zero section and $\eta$ is some curve in the base $B$.
The terms in~\eqref{eq:7} can be considered either as
elements of the homology group $H_{4}(X, {\mathbb Z})$ or, by
Poincare duality, as elements of cohomology $H^{2}(X, {\mathbb Z})$. 
This ambiguity
will occur in many of the topological expressions in this paper.

In terms of the coordinates $\bX$, $\bY$, $\bZ$ introduced above, 
it can be shown that
the spectral cover can be represented as the zero set of the polynomial
\begin{equation}
s=a_{0}\bZ^{n} + a_{2}\bX\bZ^{n-2} + a_{3}\bY\bZ^{n-3} + \ldots +
a_{n}\bX^{\frac{n}{2}} 
\label{eq:8}
\end{equation}
for $n$ even and ending in $a_{n}\bX^{\frac{n-3}{2}}\bY$ if $n$ is odd,
along with the relations~\eqref{eq:3}. This tells us that the polynomial
$s$ must be a holomorphic section of the line bundle of the spectral 
cover,
$\cO(\cC)$. That is,
\begin{equation}
s \sim \cO(n\sigma + \pi^{*}\eta)
\label{eq:9}
\end{equation}
It follows from this and equations~\eqref{eq:3} and~\eqref{eq:4}, that the
coefficients $a_{i}$ in the polynomial $s$ must be sections of the line
bundles
\begin{equation}
a_{i} \sim \pi^*K_{B}^{i} \otimes \cO(\pi^{*}\eta)
\label{eq:10}
\end{equation}
for $i=1,\ldots ,n$ where we have used expression~\eqref{eq:6}.

In order to describe vector bundles most simply, there are two properties
that we require the spectral cover to possess. The first, which is shared by
all spectral covers, is that 
\begin{itemize}
\item $\cC$ must be an effective class in $H_{4}(X,{\mathbb Z})$.
\end{itemize}
This property is simply an expression of the fact the spectral cover must be
an actual surface in $X$. It can easily be shown that 
\begin{equation}
\cC \subset X \text{ is effective } \Longleftrightarrow \eta 
\text{ is an effective class in } H_{2}(B, {\mathbb Z}).
\label{eq:12}
\end{equation}
The second property that we require for the spectral cover is that
\begin{itemize}
\item $\cC$ is an irreducible surface.
\end{itemize}
This condition is imposed because it guarantees that the associated vector
bundle is stable. It is important to note, however, that semi--stable 
vector bundles can be constructed from reducible spectral covers, 
as we will do later in this paper. Deriving the 
conditions under which $\cC$ is irreducible is not completely trivial
and will be 
discussed in detail elsewhere \cite{mathpaper}. Here, we will 
simply state the results.
First,  recall from~\eqref{eq:10} that 
$a_{i} \sim \pi^*K_{B}^{i}\otimes 
\cO(\pi^{*}\eta)$ and, hence, the zero locus of $a_{i}$ is a divisor,
$D(a_{i})$, in $X$. Then, we can show that $\cC$ is an irreducible surface if
\begin{equation}
D(a_{0}) \text{ is an irreducible divisor in } X
\label{eq:13}
\end{equation}
and
\begin{equation}
D(a_{n}) \text{ is an effective class in } H_{4}(X, {\mathbb Z}).
\label{eq:14}
\end{equation}
Using Bertini's theorem, it can be shown that condition~\eqref{eq:13} is
satisfied if the linear system $|\eta|$ is base point free. 
``Base point free'' means that for any $b \in B$, we can find a
member of the linear system $|\eta|$  that does not pass through the point $b$.

In order to make these concepts more concrete, we take, as an example, the
base surface to be
\begin{equation}
B={\mathbb F}_{r}
\label{eq:15}
\end{equation}
and derive the conditions under  
which~\eqref{eq:12},~\eqref{eq:14} and~\eqref{eq:13}
are satisfied. Recall\cite{don2} that the homology group $H_{2}({\mathbb
F}_{r}, {\mathbb Z})$ has
as a basis the effective classes $\cS$ and $\cE$ with intersection numbers
\begin{equation}
\cS^{2}=-r, \qquad \cS\cdot \cE=1, \qquad \cE^{2}=0
\label{eq:16}
\end{equation}
Then, in general, $\cC$ is given by expression~\eqref{eq:7} where
\begin{equation}
\eta= a\cS+b\cE
\label{eq:17}
\end{equation}
and $a$,$b$ are integers. One can easily check that $\eta$ is an effective
class in ${\mathbb F}_{r}$, and, hence, that $\cC$ is an effective class in
$X$, if and only if
\begin{equation}
a \geq 0, \qquad b \geq 0.
\label{eq:18}
\end{equation}
It is also not too hard to demonstrate that the linear system $|\eta|$ is base
point free if and only if
\begin{equation}
b \geq ar
\label{eq:19}
\end{equation}
Imposing this constraint then implies that $D(a_{0})$ is an irreducible divisor in
$X$. Finally, we can show that for $D(a_{n})$ to be effective in $X$ one must have
\begin{equation}
a \geq 2n, \qquad b \geq n(r+2)
\label{eq:20}
\end{equation}
We will give a detailed derivation of~\eqref{eq:19} and~\eqref{eq:20} elsewhere
\cite{mathpaper}. Combining conditions~\eqref{eq:19} 
and~\eqref{eq:20} then guarantees that
$\cC$ is an irreducible surface. To be even more specific, we now present 
two physically relevant examples of this type.

\smallskip

\

\noindent
{\bf Example 1:}
Choose the structure group of the vector bundle to be
\begin{equation}
G=SU(5)
\label{eq:21}
\end{equation}
Hence, $n=5$. Also, restrict $r=1$. Now choose
\begin{equation}
\eta= 12\cS+15\cE
\label{eq:22}
\end{equation}
so that $a=12$ and $b=15$. These parameters are easily shown to satisfy
the conditions~\eqref{eq:18},~\eqref{eq:19} and~\eqref{eq:20}. Therefore,
the associated spectral surface 
\begin{equation}
\cC=5\sigma+ \pi^{*}(12\cS+15\cE)
\label{eq:23}
\end{equation}
is both effective and irreducible.

\smallskip

\

\noindent
{\bf Example 2:}
As a second example, consider again
\begin{equation}
G=SU(5)
\label{eq:24}
\end{equation}
Hence, $n=5$. Again, restrict $r=1$. Now choose
\begin{equation}
\eta= 24\cS+36\cE
\label{eq:25}
\end{equation}
so that $a=24$ and $b=36$. These parameters also satisfy
equations~\eqref{eq:18},~\eqref{eq:19} and~\eqref{eq:20}. Therefore,
the associated 
spectral surface
\begin{equation}
\cC=5\sigma+ \pi^{*}(24\cS+36\cE)
\label{eq:26}
\end{equation}
is again both effective and irreducible.

\smallskip

\

The reason for choosing
these two examples will become clear soon. We now turn to the second
mathematical object that is required to specify an $SU(n)$ vector bundle.

\subsection*{The Line Bundle $\cN$:}

As discussed in \cite{FMW, don2}, in addition to the spectral cover 
it is necessary to 
specify a line bundle, $\cN$, over $\cC$. For $SU(n)$ vector bundles, this
line bundle must be a restriction of a global line bundle on $X$
(which we will again denote by ${\mathcal N}$), satisfying the condition
\begin{equation}
c_{1}(\cN)=n(\frac{1}{2}+\lambda)\sigma+(\frac{1}{2}-\lambda)
\pi^{*}\eta+(\frac{1}{2}+n\lambda)\pi^{*}c_{1}(B)
\label{eq:27}
\end{equation}
where $c_{1}(\cN)$, $c_{1}(B)$ are the first Chern classes of $\cN$ and $B$
respectively and $\lambda$ is, a priori, a rational number. Since $c_{1}(\cN)$
must be an integer class, it follows that either
\begin{equation}
n \quad \mbox{is odd}, \qquad \lambda=m+\frac{1}{2}
\label{eq:28}
\end{equation}
or
\begin{equation}
n \quad \mbox{is even}, \qquad \lambda=m, \qquad \eta=c_{1}(B) \quad \mod 2
\label{eq:29}
\end{equation}
where $m \in {\mathbb Z}$. In this paper, for simplicity, we will always
take examples where $n$ is odd.

\subsection*{$SU(n)$ Vector Bundle:}

Given a spectral cover, $\cC$, and a line bundle, $\cN$, satisfying the above
properties, one can now uniquely construct an $SU(n)$ vector bundle, $V$. This
can be accomplished in two ways. First, as discussed in \cite{FMW, don2}, 
the vector bundle
can be directly constructed using the associated Poincare bundle,
${\mathcal P}$.
The result is that 
\begin{equation}
V=\pi_{1*}(\pi^{*}_{2}\cN\otimes {\mathcal P})
\label{eq:30}
\end{equation}
where $\pi_{1}$ and $\pi_{2}$ are the two projections of the fiber
product $X \times_{B} \cC$ onto the two factors $X$ and $\cC$. We refer the
reader to \cite{FMW, don2} for a detailed discussion. Equivalently, $V$ can be
constructed directly from $\cC$ and $\cN$ using the Fourier-Mukai
transformation, as discussed in \cite{don3, mathpaper, AD}. 
Both of these constructions work in
reverse, yielding the spectral data $(\cC, \cN)$ up to the overall factor
of $K_B$ given the vector bundle $V$. 
Throughout this paper we will indicate this relationship between the
spectral data and the vector bundle by writing
\begin{equation}
(\cC,\cN) \longleftrightarrow V
\label{eq:31}
\end{equation}
The Chern classes for the $SU(n)$ vector bundle $V$ have been 
computed in \cite{FMW}
and \cite{don3, curio}. The results are
\begin{equation}
c_{1}(V)=0
\label{eq:32}
\end{equation}
since $\operatorname{tr} F=0$ for the structure group $SU(n)$,
\begin{equation}
c_2(V)=\eta\sigma-\frac{1}{24}c_1(B)^2(n^3-n)
+\frac{1}{2}(\lambda^2-\frac{1}{4})n\eta(\eta-nc_1(B)) 
\label{eq:33}
\end{equation}
and
\begin{equation}
c_3(V)= 2\lambda \sigma \eta(\eta-nc_1(B)). 
\label{eq:34} 
\end{equation}

In order to make these concepts more concrete, we again take the
base surface to be
\begin{equation}
B = {\mathbb F}_{r}
\label{eq:35}
\end{equation}
The Chern classes for this surface are known and given by
\begin{equation}
c_{1}({\mathbb F}_{r})=2\cS+(r+2)\cE, \qquad c_{2}({\mathbb F}_{r})=4
\label{eq:36}
\end{equation}
Now consider the two specific examples discussed above.

\smallskip

\

\noindent
{\bf Example 1:} \quad
In this example, the structure group of the vector bundle is chosen to be
$G=SU(5)$ and, hence, $n=5$. Also, we restrict $r=1$ and take
$\eta=12\cS+15\cE$. 
We further specify the line bundle, $\cN$, by choosing
\begin{equation}
\lambda=\frac{1}{2}
\label{eq:37}
\end{equation}
Note that this requirement 
is consistent with condition~\eqref{eq:28} since $n=5$ is odd.
It follows from~\eqref{eq:16},~\eqref{eq:33} and~\eqref{eq:36} that
\begin{equation}
c_{2}(V)= (12\cS+15\cE)\sigma-40F
\label{eq:38}
\end{equation}
where $F$ is the generic class of the fiber,
and from~\eqref{eq:16},~\eqref{eq:34} and~\eqref{eq:36} that
\begin{equation}
c_{3}(V)= 6
\label{eq:39}
\end{equation}

\smallskip

\

\noindent
{\bf Example 2:} \quad
As a second example, consider
again $G=SU(5)$, $n=5$, $r=1$ and we take $\eta= 24S+36\cE$.
Further specify the line bundle, $\cN$, by choosing
\begin{equation}
\lambda=\frac{1}{2}
\label{eq:40}
\end{equation}
It follows from~\eqref{eq:16},~\eqref{eq:33} and~\eqref{eq:36} that
\begin{equation}
c_{2}(V)= (24\cS+36\cE)\sigma-40F
\label{eq:41}
\end{equation}
and from~\eqref{eq:16},~\eqref{eq:34} and~\eqref{eq:36} that
\begin{equation}
c_{3}(V)= 672
\label{eq:42}
\end{equation}

\smallskip

\

To conclude, in this section we have discussed the construction and properties
of stable $SU(n)$ vector bundles associated with irreducible spectral covers.
For the remainder of this paper, for brevity, we will refer to such bundles
simply as ``stable $SU(n)$ vector bundles''.


\section{Physical Topological Conditions:}


As discussed in a number of papers \cite{nse, curio, lpt, GSW}, 
there are two fundamental conditions
that must be satisfied in any physically acceptable heterotic M--theory. 

\subsection*{Anomaly Cancellation:}

The first condition is a direct consequence of demanding that the theory be
anomaly free. This requires that the Bianchi identity for the field strength
of the three--form be modified by the addition of both gauge and tangent
bundle Chern classes as well as by sources for possible bulk space
five--branes. Integrating this modified Bianchi identity over an
arbitrary four--cycle then gives the topological condition
\[
c_{2}(V_{1})+c_{2}(V_{2})+W=c_{2}(TX)
\]
where $V_{1}$ and $V_{2}$ are the vector bundles on the observable and hidden
boundary branes respectively and $TX$ is the tangent bundle of the Calabi--Yau
threefold $X$. In addition, $W$ is the class of the holomorphic 
curve in $X$ around which possible bulk space five--branes are
wrapped. Any anomaly 
free heterotic M--theory must satisfy this condition. In this paper, for
simplicity, we will always choose the vector bundle on the hidden sector,
$V_{2}$, to be trivial. This ensures an unbroken $E_{8}$ gauge group in the
hidden sector and simplifies equation~\eqref{eq:43} to
\begin{equation}
W=c_{2}(TX)-c_{2}(V)
\label{eq:43}
\end{equation}
where we now denote $V_{1}$ by $V$. Given the second Chern classes for 
$V$ and $TX$, this equation acts as a definition for the five--brane class
$W$. As such, it is not, a priori, a constraint.
However, the five--brane class must, on physical grounds, be represented by an
 actual surface in $X$. Hence, 
\begin{equation}
W  \text{ must be an effective class in } H_{2}(X, {\mathbb Z}).
\label{eq:44}
\end{equation}
This condition puts a non--trivial constraint on the choice of the vector
bundle $V$. It is important to note, however, that the constraint of
effectiveness of $W$ is much less restrictive then trying to set
$c_{2}(V)=c_{2}(TX)$, as we will see.

Using equation~\eqref{eq:33} and the fact that
\begin{equation}
c_{2}(TX)=(c_{2}(B)+11c_{1}(B)^{2})F +12\sigma\cdot c_{1}(B)
\label{eq:45}
\end{equation}
where $c_{2}(B)$ is the second Chern class of the base $B$, 
it follows from~\eqref{eq:43} that
\begin{equation}
W=W_{B}\sigma +a_{f}F
\label{eq:46}
\end{equation}
where
\begin{equation}
W_{B}=\pi^{*}(12c_{1}(B)-\eta)
\label{eq:47}
\end{equation}
and
\begin{equation}
a_{f}=c_{2}(B)+(11+\frac{n(n^{2}-1)}{24})c_{1}(B)^{2}-\frac{n}{2}
(\lambda^{2}-\frac{1}{4})\eta(\eta-nc_{1}(B)).
\label{eq:48}
\end{equation}
For concreteness, let us evaluate these quantities for the two sample cases
discussed above.

\smallskip

\

\noindent
{\bf Example 1:} \quad In this case, $G=SU(n)$, $n=5$, $B={\mathbb F}_{1}$, 
$\eta=12\cS+15\cE$ and
$\lambda=\frac{1}{2}$. It then follows
from~\eqref{eq:16},~\eqref{eq:36},~\eqref{eq:47} and~\eqref{eq:48} that 
\begin{equation}
W_{B}=12\cS+21\cE, \qquad a_{f}=132
\label{eq:49}
\end{equation}

\smallskip

\

\noindent
{\bf Example 2:} \quad Here $G=SU(n)$, $n=5$, $B={\mathbb F}_{1}$, 
$\eta=24\cS+36\cE$ and
$\lambda=\frac{1}{2}$. It then follows
from~\eqref{eq:16},~\eqref{eq:36},~\eqref{eq:47} and~\eqref{eq:48} that 
\begin{equation}
W_{B}=0, \qquad a_{f}=132
\label{eq:50}
\end{equation}

\smallskip

\

These examples elucidate two important properties of five--brane classes
$W$ within the context of the stable $SU(n)$ vector bundles discussed so
far. The first property is that it is possible to choose vector bundles such
that $W_{B}=0$, as in Example 2. We will show in the next section that, 
under certain conditions, a vacuum with a five--brane 
class with non-vanishing base component, $W_{B} \neq 0$, can make a phase
transition via a ``small instanton'' to a vacuum in which the five--brane 
class base component vanishes, $W_{B}=0$. Note, however, that 
in both of the above examples the five--brane
fiber component, $a_{f}F$, is non-zero. This turns out to indicate the second
property of five--brane classes. That is, it is never possible, 
within context of the stable $SU(n)$ vector bundles arising from 
irreducible spectral surfaces,  
to choose a vector bundle such that the entire five--brane class vanishes,
$W=0$. This statement is sufficiently important for us to provide a short
proof.

\subsection*{Search for $c_{2}(V)=c_{2}(TX)$:}

In order for $W=0$, it is necessary that $W_{B}$ in~\eqref{eq:47} and
$a_{f}$ in~\eqref{eq:48} both vanish. Clearly, $W_{B}=0$ requires that
one choose 
\begin{equation}
\eta=12c_{1}(B)
\label{eq:51}
\end{equation}
Inserting this expression into~\eqref{eq:48}, we find that $a_{f}$
will vanish if 
and only if
\begin{equation}
\lambda=\pm\sqrt{\frac{c_{2}+(11+\frac{n(n^{2}-1)}{24}
+\frac{3n(12-n)}{2})c_{1}^{2}}{6n(12-n)}}
\label{eq:52}
\end{equation}
Recall that for $c_{1}(\cN)$ to be an integer class, the parameter $\lambda$ must
be a rational number. The square root of a rational number is generically not
rational itself and, hence, we do not expect $\lambda$ in~\eqref{eq:52} to be
rational. We have checked for del Pezzo, Hirzebruch and Enriques bases with
physically sensible value of $n$ and found that this is indeed the case. We
conclude, therefore, that one can never have $W=0$ or, equivalently,
that one can never set $c_{2}(V)=c_{2}(TX)$ within context of the 
stable $SU(n)$ vector bundles discussed so far.

\subsection*{Number of Generations:}

The second fundamental topological condition is the statement that the number
of families of quarks and leptons on the observable brane is given by
\[
N_{gen}=\frac{c_{3}(V)}{2}
\]
It follows from~\eqref{eq:34} that
\begin{equation}
N_{gen}=\lambda \sigma \eta( \eta-nc_{1}(B))
\label{eq:53}
\end{equation}
For concreteness, let us evaluate the number of generations for the two 
sample cases discussed above.

\smallskip

\

\noindent
{\bf Example 1:} \quad 
In this case, $G=SU(n)$, $n=5$, $B={\mathbb F}_{1}$, $\eta=12\cS+15\cE$ and
$\lambda=\frac{1}{2}$. It then follows
from~\eqref{eq:16},~\eqref{eq:36} and~\eqref{eq:53} that 
\begin{equation}
N_{gen}=3
\label{eq:54}
\end{equation}

\smallskip

\

\noindent
{\bf Example 2:} \quad In this example 
$G=SU(n)$, $n=5$, $B={\mathbb F}_{1}$, $\eta=24\cS+36\cE$ and
$\lambda=\frac{1}{2}$. It then follows
from~\eqref{eq:16},~\eqref{eq:36} and~\eqref{eq:53} that 
\begin{equation}
N_{gen}=336
\label{eq:55}
\end{equation}

\smallskip

\

Once again, these examples are indicative of an important property of phase
transitions via ``small instantons'' that change a five--brane class with
$W_{B} \neq 0$ into a five--brane class with $W_{B}=0$. That is, in any such
transition the number of generations will change. Thus, such processes
correspond to ``chirality--changing'' phase transitions.


\section{Chirality--Changing Small Instanton Transitions:}
\label{sec-chirality}


In this section, we will show that, within the 
context of the stable $SU(n)$ vector bundles discussed so far, 
chirality--changing phase transitions via small
instantons can occur. These transitions have the property that they 
take a vacuum with a non-zero base
component, $W_{B} \neq 0$, in the five--brane class and transform it to a
vacuum with a five--brane class with either a vanishing base component,
$W_{B}=0$, or a smaller base component.
Such transitions do not affect the fiber component, $a_{f}F$, of the
five--brane curve, which is non-vanishing and identical on both sides of the
small instanton transition. Phase transitions which change the fiber component
of $W$ will be discussed in the next section.

Let us begin on the observable boundary brane 
with a stable  $SU(n)$ vector bundle $V$ specified by the
spectral cover $\cC$ which is a smooth, irreducible surface in the
homology class 
\begin{equation}
{\cal{C}}=n\sigma + \pi^{*}\eta
\label{eq:56}
\end{equation}
and the line bundle $\cN$ over ${\cal{C}}$ whose first Chern class
satisfies~\eqref{eq:27}. It follows from the above discussion that the bulk
space five--brane class $W=W_{B}\sigma+a_{f}F$ does not vanish. 
In addition, we will demand that $V$ and $TX$ be such that the base component
\begin{equation}
W_{B} \neq 0
\label{eq:57}
\end{equation}
The fiber component, $a_{f}F$, also does not vanish, but will not concern us in
this section. Since any $W$ in $X$ is a four--form, it follows from Poincare
duality that $W_{B}$ must be a surface in $X$. From~\eqref{eq:47}, we see that 
$W_{B}$ is of the form $W_{B} = \pi^{*}z$ where $z=12c_{1}(B) - \eta$.

Let us now move the bulk five--brane to the observable 
boundary brane and attempt to ``absorb'' the $\pi^{*}z$ part of the 
five--brane class into the vector bundle. A full chirality--changing phase
transition will occur if we choose $z=12c_{1}(B) - \eta$. However, a
``partial'' transition can occur when $z$ is any effective subcurve that
splits off from $12c_{1}(B) - \eta$. In either case, this will result in a
``bundle'', $\widetilde{V}$, whose spectral cover, $\widetilde{\cC}$, is
reducible and of the form
\begin{equation}
\widetilde{\cC}= \cC\cup \pi^{*}z
\label{eq:59}
\end{equation}
One might try to specify the line bundle $\widetilde{\cN}$ over
$\widetilde{\cC}$ directly and then to construct the ``bundle'' $\widetilde{V}$ from
this spectral data via the Fourier--Mukai transformation. However, we find it
expedient to first discuss the properties of $\widetilde{V}$
and then use these properties to derive the general form of $\widetilde{\cN}$.
To do this, we begin by employing 
the Fourier--Mukai transformation to construct $V$ from the spectral data
$(\cC, \cN)$. That is,
\begin{equation}
({\cal{C}}, \cN) \longrightarrow V
\label{eq:60}
\end{equation}
where $V$ is the original rank $n$ vector bundle over $X$.
Second, we must specify a line bundle $\ell$ on the surface
$\pi^{*}z$ which we will glue together with $\cN$ to produce
$\widetilde{\cN}$. A
priori, $\ell$ is arbitrary, but, as we will see, it is actually subject to
strong constraints. To find these constraints, we first use the Fourier--Mukai
transformation to construct a vector bundle $V_{z}$ from the spectral data 
$(\pi^{*}z,\ell)$. That is,
\begin{equation}
(\pi^{*}z, \ell) \longrightarrow V_{z}
\label{eq:61}
\end{equation}
where $V_{z}$ is a rank $1$ vector bundle over the curve $\tilde{z} =
\sigma\cdot \pi^{*}z$.\footnote{As a technical aside, we note that the vector bundle
$V_{z}$ over the curve $\tilde{z}$ can be formally extended over the Calabi--Yau
threefold $X$ as an object that vanishes everywhere outside the curve
$\tilde{z}$ and is identical to $V_{z}$ on $\tilde{z}$. 
This extended object will also be
denoted by $V_{z}$. We will let context dictate which of these notions is to
be used. This remark applies to all of the line bundles discussed throughout
this paper.} Now, the fact that, by construction, the base component of the
bulk five--brane class associated with $\widetilde{V}$ must be equal to $W_{B} -
\pi^{*}z$, implies that $\ell = \pi^{*}L$ for some line bundle $L$ on the
curve $z$ in the base. A simple Fourier-Mukai calculation shows that
$V_{z}$ is  just
\begin{equation}
V_{z}=i_{*}(L \otimes K_{B})
\label{eq:62}
\end{equation}
where $i$ is the embedding $i : \tilde{z} \rightarrow X$ of the
curve $\tilde{z}$ in $X$.

Given $V$ and $V_{z}$, we now attempt to ``weave'' them
together by relating them on the curve $\tilde{z}$ in $X$ where their
data overlap. This is done by specifying a
surjection
\begin{equation}
\xi:V|_{\tilde{z}} \longrightarrow V_{z}
\label{eq:63}
\end{equation}
That such a surjection exists and is unique will become clear shortly. Given
this relation, one can define a ``bundle'' $\widetilde{V}$ on $X$ via the
exact sequence\footnote{The bundle $\widetilde{V}$ defined in this 
way bears a special name. It is a called a Hecke transform of $V$ and the pair 
$(\xi, V_{z})$ is called the center of the Hecke transform.}
\begin{equation}
0 \rightarrow \widetilde{V} \rightarrow V \rightarrow V_{z}
\rightarrow 0
\label{eq:64}
\end{equation}
It can be shown that because
\begin{equation}
\operatorname{codimension}   \tilde{z}= 2 > 1,
\label{eq:65}
\end{equation}
then $\tilde{V}$ is a singular object, 
called a torsion free sheaf\footnote{The notion of stability here is similar
to that used for vector bundles. The differential geometric counterpart of
a stable sheaf $\widetilde{V}$ is a Hermitian-Yang-Mills
connection $\widetilde{A}$ on the vector
bundle $V$ which is smooth outside of the curve $\tilde{z} \subset X$
but has a delta function behavior along the curve $\widetilde{z}$. In
other words, a stable torsion free sheaf $\widetilde{V}$ is the
algebraic-geometry incarnation of a ``small instanton''
concentrated on the curve $\widetilde{z}$. This justifies the terminology
``small instanton phase transition''that we use below to describe the
two step process of first creating $\widetilde{V}$ out of $V$ and then
deforming $\widetilde{V}$ to a smooth vector bundle $\widehat{V}$ on
$X$.}, but is not a smooth
vector bundle. Hence, to complete our construction, we will have to show that
$\widetilde{V}$ can be ``smoothed out'' to a stable vector bundle. Before
doing this, however, we must first compute the Chern classes of the
torsion free sheaf $\widetilde{V}$. It follows from the exact
sequence~\eqref{eq:64} that
\begin{equation}
Ch(\widetilde{V})=Ch(V) + Ch(V_{z})
\label{eq:66}
\end{equation}
where $Ch$ stands for the Chern character. Using the
Grothendieck--Riemann--Roch theorem, we find that
\begin{equation}
Ch(V_{z})=i_{*}(Ch(L\otimes K_{B|z})Td(z))Td(X)^{-1}
\label{eq:67}
\end{equation}
where $Td$ stands for the Todd class. Inserting this expression
into~\eqref{eq:66} and expanding out, one obtains the Chern classes
\begin{equation}
c_{1}(\widetilde{V})=c_{1}(V)=0,
\label{eq:68}
\end{equation}
\begin{equation}
c_{2}(\tilde{V})=c_{2}(V) +z
\label{eq:69}
\end{equation}
\begin{equation}
c_{3}(\tilde{V})=c_{3}(V)-2(c_{1}(V_{z})+1-g)
\label{eq:70}
\end{equation}
where $c_{1}(V)$,$c_{2}(V)$ and $c_{3}(V)$ are the Chern classes of the bundle
$V$ given in~\eqref{eq:32},~\eqref{eq:33} and~\eqref{eq:34} respectively
and $g$ is the genus of the curve $z$.
It follows from~\eqref{eq:62} that
\begin{equation}
c_{1}(V_{z})=c_{1}(L) -c_{1}(B) \cdot \tilde{z}
\label{eq:71A}
\end{equation}
Using the Riemann--Roch formula on $B$, one can also show that the
genus is given by 
\begin{equation}
1-g=\frac{1}{2}(c_{1}(B)-z)\cdot z
\label{eq:71B}
\end{equation}
Inserting these expressions into~\eqref{eq:70}, the third Chern
class of $\widetilde{V}$ can be written as
\begin{equation}
c_{3}(\widetilde{V})=c_{3}(V)-2c_{1}(L) +(c_{1}(B)+z)\cdot z
\label{eq:71C}
\end{equation}
Note that this result is true for any allowed choice of the line bundles
$\cN$ on $\cC$ and $L$ on $z$. Having defined the torsion free sheaf 
$\widetilde{V}$, we can now construct its spectral line bundle 
$\widetilde{\cN}$ via the
inverse Fourier--Mukai transformation
\begin{equation}
\widetilde{V} \longrightarrow (\widetilde{\cC},\widetilde{\cN})
\label{eq:72}
\end{equation}
where $\widetilde{\cC}$ is given in~\eqref{eq:59}. It follows
from the Fourier--Mukai transformation and~\eqref{eq:64} 
that $\widetilde{\cN}$ must lie in the exact sequence
\begin{equation}
0 \rightarrow \pi^{*}L \rightarrow \widetilde{\cN} \rightarrow 
\cN \rightarrow 0,
\label{eq:73}
\end{equation}
where $\cN$ and $\widetilde{\cN}$ are understood as line bundles on
$\cC$ and $\widetilde{\cC}$ respectively.

This sequence implies that the line bundles $\widetilde{\cN}$ and
$\pi^{*}L$  on $\pi^{*}z$ are not 
independent but, rather, are related by the expression
\begin{equation}
\left(\widetilde{\cN} \otimes \cO_{X}(-(\cC \cdot \pi^{*}z))
\right)|_{\pi^{*}z}=\pi^{*}L
\label{eq:74}
\end{equation}
Evaluating the first Chern class on $\pi^{*}z$, we find that
\begin{equation}
c_{1}(\widetilde{\cN})|_{\pi^{*}z} -\cC \cdot \pi^{*}z= \pi^{*}c_{1}(L)
\label{eq:75}
\end{equation}
We will return to this equation and a
discussion of the properties of $\widetilde{\cN}$ below.

We can now examine whether the singular torsion free sheaf, $\widetilde{V}$, 
can be smoothed out to a stable $SU(n)$ vector bundle, 
which we will denote by $\widehat{V}$. The spectral data of  
$\widehat{V}$ can be
obtained via the inverse Fourier--Mukai transformation
\begin{equation}
\widehat{V} \longrightarrow (\widehat{\cC}, \widehat{\cN})
\label{eq:76}
\end{equation}
Clearly, the spectral cover $\widehat{C}$ is in the homology class
\begin{equation}
\widehat{\cC} = n\sigma +\pi^{*}\hat{\eta}
\label{eq:77}
\end{equation}
where
\begin{equation}
\hat{\eta}=\eta +z.
\label{eq:78}
\end{equation}
To ensure that $\widehat{V}$ is stable we need to take $\widehat{\cC}$
to be an irreducible surface in the class $n\sigma +
\pi^{*}\hat{\eta}$. 
Furthermore, from~\eqref{eq:27} we see that the line bundle
$\widehat{\cN}$ must satisfy 
\begin{equation}
c_{1}(\widehat{\cN})=n(\frac{1}{2}+\lambda)\sigma+(\frac{1}{2}-\lambda)
\pi^{*}\hat{\eta}+(\frac{1}{2}+n\lambda)\pi^{*}c_{1}(B)
\label{eq:79}
\end{equation}
The structure of $\widehat{\cN}$ will be further discussed below.

Since $\widehat{V}$ is a smooth, stable $SU(n)$ bundle, 
then it follows from~\eqref{eq:32},~\eqref{eq:33} and~\eqref{eq:34} that
\begin{equation}
c_{1}(\widehat{V})=0,
\label{eq:80}
\end{equation}
\begin{equation}
c_2(\widehat{V})=\hat{\eta}\sigma-\frac{1}{24}c_1(B)^2(n^3-n)
+\frac{1}{2}(\lambda^2-\frac{1}{4})n\hat{\eta}(\hat{\eta}-nc_1(B)), 
\label{eq:81}
\end{equation}
\begin{equation}
c_3(\widehat{V})= 2\lambda \sigma \hat{\eta}(\hat{\eta}-nc_1(B)). 
\label{eq:82} 
\end{equation}
If the bundle $\widehat{V}$ exists, then its Chern classes must match
those of the 
torsion free sheaf $\widetilde{V}$. That is, we must have
\begin{equation}
c_{i}(\widehat{V})=c_{i}(\widetilde{V})
\label{eq:82A}
\end{equation}
for $i=1,2,3$. It follows from~\eqref{eq:80}
and~\eqref{eq:68} that both first Chern classes vanish. Comparing the second
Chern classes given in~\eqref{eq:81} and~\eqref{eq:69} respectively, 
we see that they will be identical if and only if one restricts the spectral
line bundle $\cN$ of the original 
bundle $V$ so that
\begin{equation}
\lambda=\pm\frac{1}{2}
\label{eq:83}
\end{equation}
Hence, small instanton transitions of this type only occur for certain
components of the moduli space of 
$SU(n)$ vector bundles. Henceforth, we will assume
that~\eqref{eq:83} is satisfied. Inserting these values for $\lambda$ into 
expression~\eqref{eq:82} for the third Chern class of $\widehat{V}$, we find,
using~\eqref{eq:78}, that it will be identical to $c_{3}(\widetilde{V})$
in~\eqref{eq:71C} if and only if
\begin{equation}
c_{1}(L)=\left(\frac{1}{2}(1\pm n)c_{1}(B)+\frac{1}{2}(1\mp 1)z 
\mp\eta\right) \cdot z 
\label{eq:84}
\end{equation}
Therefore, the line bundle $L$ on the
 curve $z$ is also not arbitrary, but must
satisfy constraint~\eqref{eq:84}. Note that, in general, $d \in
 {\mathbb Z}$ but need not be positive.

If the torsion free sheaf $\widetilde{V}$ can be smoothed out to the 
irreducible vector bundle ${\widehat{V}}$, then, 
in addition to~\eqref{eq:82A}, 
the corresponding spectral line bundles must satisfy
\begin{equation}
c_{1}(\widehat{\cN})=c_{1}(\widetilde{\cN})
\label{eq:85}
\end{equation}
Inserting this into expression~\eqref{eq:75}, we have
\begin{equation}
c_{1}(\widehat{\cN})|_{\pi^{*}z} -\cC \cdot \pi^{*}z= \pi^{*}c_{1}(L)
\label{eq:85A}
\end{equation}
Using~\eqref{eq:56} and~\eqref{eq:79}, we find that 
\begin{equation}
n(\lambda-\frac{1}{2})\sigma \cdot \pi^{*}z +\pi^{*}\left(((\frac{1}{2}-
\lambda)(\eta+z)+ (\frac{1}{2}+ n\lambda)c_{1}(B)-\eta) \cdot z \right)=
\pi^{*}c_{1}(L)
\label{eq:86}
\end{equation}
Since the first term on the left hand side is not of the form $\pi^{*}$ of
some expression, it follows that we must take
\begin{equation}
\lambda= \frac{1}{2}
\label{eq:87}
\end{equation}
for consistency. This is compatible with~\eqref{eq:83}, but is a stronger
constraint. Henceforth, we will assume that~\eqref{eq:87} is satisfied. 
In this case, expression~\eqref{eq:86} simplifies to
\begin{equation}
c_{1}(L)=(\frac{1}{2}(1+n)c_{1}(B)-\eta) \cdot z
\label{eq:87A}
\end{equation}
which is compatible with~\eqref{eq:84} for the choice of
$\lambda=\frac{1}{2}$. It follows that the spectral line bundle $L$ on the
curve $z$ is not only not arbitrary, but is uniquely fixed to be
\begin{equation}
L=(\frac{1}{2}(1+n)c_{1}(B)-\eta) \cdot z
\label{eq:87B}
\end{equation}
Furthermore, note that for $\lambda= \frac{1}{2}$
\begin{equation}
c_{1}(\widehat{\cN})=c_{1}(\cN)
\label{eq:88}
\end{equation}
and, therefore   , using~\eqref{eq:85} that
\begin{equation}
\widehat{\cN}=\widetilde{\cN}=\cN
\label{eq:89}
\end{equation}
It follows that the spectral line bundles $\widehat{\cN}$ and
$\widetilde{\cN}$  are
also uniquely fixed in terms of $\cN$. 

We conclude that for the choice of $\lambda=\frac{1}{2}$ and $L$ given
by~\eqref{eq:87B}, the torsion free sheaf $\widetilde{V}$ can be smoothed out to
a stable $SU(n)$ vector bundle $\widehat{V}$ and, hence, the phase
transition can be completed. Note from~\eqref{eq:71C},~\eqref{eq:82A}
and~\eqref{eq:87A} that
\begin{equation}
c_{3}(\widehat{V})=c_{3}(V)+ (2\eta+z-nc_{1}(B))\cdot z
\label{eq:90}
\end{equation}
It follows from~\eqref{eq:53} that such phase transitions generically change
the number of generations.

\subsection*{Summary:}

In this section we have shown the following.

\begin{itemize}

\item Start with a heterotic M--theory vacuum specified by
a stable $SU(n)$ vector bundle, $V$, on the observable boundary
brane with spectral cover
\begin{equation}
{\cal{C}}= n\sigma+ \pi^{*}\eta
\label{eq:91}
\end{equation}
and line bundle $\cN$ constrained to have
\begin{equation}
\lambda=\frac{1}{2}
\label{eq:92}
\end{equation}
as well as a five--brane class in the bulk space 
\begin{equation}
W=W_{B}\sigma +a_{f}F
\label{eq:93}
\end{equation}
where $W_{B}$ is non-vanishing. Note that
\begin{equation}
W_{B}=\pi^{*}z
\label{eq:94}
\end{equation}
where $z=12c_{1}(B)-\eta$.

\item Now move the five--brane through the bulk space until it touches 
the observable brane and ``detach'' either
all, or a portion, of the base component of the five--brane class.
That is, consider $\pi^{*}z$, where $z$ is either the entire base curve
$12c_{1}(B)-\eta$ or some effective subcurve.
Leave the rest of the base component, if any, and the pure fiber component, 
$a_{f}F$, of the five--brane class undisturbed. One can then define a 
rank $1$ vector bundle $V_{z}$ over the curve $\tilde{z}=\sigma \cdot
\pi^{*}z$ with spectral data $(\pi^{*}z,\pi^{*}L)$
where
\begin{equation}
L=(\frac{1}{2}(1+n)c_{1}(B)-\eta) \cdot z
\label{eq:95}
\end{equation}

\item The original vector bundle, $V$, now combines with the rank $1$ vector
bundle, $V_{z}$, to form a singular torsion free sheaf, 
$\widetilde{V}$, on the observable
brane. This sheaf has a reducible spectral cover
\begin{equation}
\widetilde{\cC}={\cal{C}} \cup \pi^{*}z
\label{eq:96}
\end{equation}
and spectral line bundle
\begin{equation}
\widetilde{\cN}=\cN
\label{eq:97}
\end{equation}
This singular torsion free sheaf is called a small instanton.

\item The small instanton can now be smoothed out into a stable $SU(n)$
vector bundle $\widehat{V}$ with spectral cover
\begin{equation}
\widehat{\cC}= n\sigma+ \pi^{*}(\eta+z)
\label{eq:98}
\end{equation}
and spectral line bundle
\begin{equation}
\widehat{\cN}=\cN
\label{eq:99}
\end{equation}
Note that $\widehat{V}$ has the same structure group, $SU(n)$, as $V$ and that both
$\widehat{\cN}$ and $\cN$ have $\lambda=\frac{1}{2}$.

\item The Chern classes of the original vector bundle $V$ and the final 
vector bundle $\widehat{V}$ after the phase transition are related by
\begin{equation}
c_{1}(\widehat{V})=c_{1}(V)=0,
\label{eq:100}
\end{equation}
\begin{equation}
c_{2}(\widehat{V})=c_{2}(V) +z
\label{eq:101}
\end{equation}
\begin{equation}
c_{3}(\widehat{V})=c_{3}(V)+ (2\eta+z-nc_{1}(B))\cdot z
\label{eq:102}
\end{equation}

\end{itemize}

These operations define a chirality--changing small instanton phase transition
from one heterotic M-theory vacuum to another involving either all, or part, of
the base component of the five--brane class. The remainder the base
component, if any, and the entire pure fiber class have not been involved in
this transition. In order to make these concepts more transparent, we now
present several examples.

\bigskip

\noindent
{\bf Example 1:}
Consider the first sample vacuum discussed earlier in this paper, specified by
$B={\mathbb F}_{1}$, $G=SU(5)$, spectral cover
\begin{equation}
\cC=5\sigma+\pi^{*}\eta
\label{eq:103}
\end{equation}
where
\begin{equation}
\eta=12\cS+15\cE
\label{eq:104}
\end{equation}
and line bundle $\cN$ with
\begin{equation}
\lambda=\frac{1}{2}
\label{eq:105}
\end{equation}
We found in~\eqref{eq:49} that
\begin{equation}
W_{B}=12\cS+21\cE, \qquad a_{f}=132
\label{eq:106}
\end{equation}
and in~\eqref{eq:54} that 
\begin{equation}
N_{gen}=3
\label{eq:107}
\end{equation}
Since $\lambda=\frac{1}{2}$ and $W_{B} \neq 0$, this vacuum satisfies the
criteria to make a chirality--changing small instanton transition. To specify
this, we must choose the portion of base component $W_{B}$ we wish to
``absorb'' during the transition. Let us choose the entire base curve
\begin{equation}
z=12\cS+21\cE
\label{eq:108}
\end{equation}
In this case, the small instanton transition will be to a new, irreducible
vacuum specified by $B={\mathbb F}_{1}$, $G=SU(5)$, spectral curve
\begin{equation}
\widehat{\cC}=5\sigma+\pi^{*}(\eta+ z)
\label{eq:109}
\end{equation}
where
\begin{equation}
\eta+ z=24\cS+36\cE
\label{eq:110}
\end{equation}
and line bundle $\widehat{\cN}=\cN$ and, hence,
\begin{equation}
\lambda=\frac{1}{2}
\label{eq:111}
\end{equation}
Since we have absorbed the entire base component $W_{B}$, we must have
\begin{equation}
\hat{W}_{B}=0
\label{eq:112}
\end{equation}
On the other hand, the pure fiber component of the five--brane class is left
undisturbed and, therefore,
\begin{equation}
\hat{a}_{f}=132
\label{eq:113}
\end{equation}
We note from~\eqref{eq:26},~\eqref{eq:40} and~\eqref{eq:50} 
that these are exactly the properties of the second sample vacuum that
we discussed above. Furthermore, inserting
$n=5$,~\eqref{eq:104},~\eqref{eq:107} and~\eqref{eq:108} into~\eqref{eq:102} we
find, using $N_{gen}=\frac{1}{2}c_{3}$, that
\begin{equation}
\hat{N}_{gen}=336
\label{eq:114}
\end{equation}
which is consistent with the result given in~\eqref{eq:55}. We conclude that
the two sample vacua discussed above are related to each other by a
chirality--changing small instanton transition in which the entire base
component of the five--brane class of the first vacuum is ``absorbed''.

\bigskip

\noindent
{\bf Example 2:} \quad
As a second example, consider a heterotic M--theory vacuum specified by
$B={\mathbb F}_{0}$, $G=SU(3)$, spectral cover
\begin{equation}
{\cal{C}}=3\sigma+\pi^{*}\eta
\label{eq:115}
\end{equation}
where
\begin{equation}
\eta=6\cS+6\cE
\label{eq:116}
\end{equation}
and line bundle $\cN$ with
\begin{equation}
\lambda=\frac{1}{2}
\label{eq:117}
\end{equation}
Using the formalism presented in Sections 2 and 3, it can easily be shown that
\begin{equation}
W_{B}=18\cS+18\cE, \qquad a_{f}=100
\label{eq:118}
\end{equation}
and that 
\begin{equation}
N_{gen}=0
\label{eq:119}
\end{equation}
Since $\lambda=\frac{1}{2}$ and $W_{B} \neq 0$, this vacuum satisfies the
criteria to make a chirality--changing small instanton transition. To specify
this, we must choose the portion of base component $W_{B}$ we wish to
``absorb'' during the transition. In this example, we will only choose an
effective subcurve of the base component
\begin{equation}
z=\cE
\label{eq:120}
\end{equation}
For this case, the small instanton transition will be to a new, irreducible
vacuum specified by $B={\mathbb F}_{0}$, $G=SU(3)$, spectral curve
\begin{equation}
\widehat{\cC}=3\sigma+\pi^{*}(\eta+ z)
\label{eq:121}
\end{equation}
where
\begin{equation}
\eta+ z=6\cS+7\cE
\label{eq:122}
\end{equation}
and line bundle $\widehat{\cN}=\cN$ and, hence,
\begin{equation}
\lambda=\frac{1}{2}
\label{eq:123}
\end{equation}
Since we have absorbed $\pi^{*}z$ where $z=\cE$, we must have
\begin{equation}
\widehat{W}_{B}=18\cS+17\cE
\label{eq:124}
\end{equation}
On the other hand, the pure fiber component of the five--brane class is left
undisturbed and, therefore,
\begin{equation}
\hat{a_{f}}=100
\label{eq:125}
\end{equation}
Using~\eqref{eq:102} and the data presented in this example, we can compute
the number of generations in the new vacuum after the phase transition. We
find that
\begin{equation}
\hat{N}_{gen}=3
\label{eq:126}
\end{equation}
We conclude that, in this example, we have a chirality--changing small
instanton transition from a vacuum with no families to a vacuum with three
families. The three family vacuum has both non-vanishing 
base and fiber components of its five--brane class.


\section{Gauge Group Changing Small Instanton Transitions:}


In the previous section, we discussed chirality--changing 
phase transitions involving all, or a portion, of the base component, 
$W_{B}$, of the five--brane class, $W$. This discussion did not include the pure
fiber component, $a_{f}F$, of $W$, which was left ``unabsorbed'' by the
transition. In this section, we turn our attention to the pure fiber
component. We show that all, or a portion, of $a_{f}F$ 
can always be ``absorbed'' via a small instanton
phase transition into a smooth vector bundle on the observable brane. This
vector bundle, however, is somewhat different than the stable
$SU(n)$ bundles discussed previously. Among other properties, it is a 
reducible, semi--stable vector bundle and has the product 
structure group $SU(n) \times SU(m)$. 
It follows that pure fiber component small instanton transitions generically 
change the structure group and, hence, the gauge group on the observable
brane.

\subsection*{Reducible Vector Bundles:}

We begin by considering a stable $SU(n)$ vector bundle,
$\overline{V}$, over 
$X$ specified by a spectral cover 
\begin{equation}
\bar{\cal{C}}=n\sigma+\pi^{*}\bar{\eta}
\label{eq:127}
\end{equation}
and line bundle $\overline{\cN}$ over $\bar{\cal{C}}$. The Chern classes
$c_{1}(\overline{\cN})$ and $c_{i}(\overline{V})$ for $i=1,2,3$ are given in
with $\eta$ replaced by $\bar{\eta}$.

Next, let $M$ be a stable $SU(m)$ vector bundle with $m\geq2$, 
not over $X$, but over the base $B$. The Chern classes of $M$ are trivial to
compute and are given by
\begin{equation}
c_{1}(M)=0
\label{eq:128}
\end{equation}
\begin{equation}
c_{2}(M)=k, \qquad k\in {\mathbb Z}
\label{eq:129}
\end{equation}
\begin{equation}
c_{i}(M)=0, \qquad i\geq3
\label{eq:130}
\end{equation}
Now consider the pull--back to a stable $SU(m)$ vector bundle 
$\pi^{*}M$ over $X$. The spectral data can be determined by performing an
inverse Fourier--Mukai transformation. The result is that
\begin{equation}
\pi^{*}M \longrightarrow (m\sigma, M)
\label{eq:131}
\end{equation}
where the spectral cover $m\sigma$ consists of $m$ coincident sections, called
a non-reduced surface \cite{BJPS}. Although of rank $m$, 
$\pi^{*}M$ splits into a direct
sum of $m$ one--dimensional spaces over each point in the base $B$. Hence, $M$
can be thought of as a deformation of 
a line bundle over $m\sigma$. The Chern classes of
$\pi^{*}M$ follow directly from the above and are given by
\begin{equation}
c_{1}(\pi^{*}M)=0
\label{eq:132}
\end{equation}
\begin{equation}
c_{2}(\pi^{*}M)=kF, \qquad k \in {\mathbb Z}
\label{eq:133}
\end{equation}
\begin{equation}
c_{3}(\pi^{*}M)=0
\label{eq:134}
\end{equation}

Having presented the two stable vector bundles $\overline{V}$ and
$\pi^{*}M$, 
we now construct a reducible bundle, $\bV$, by taking their
direct sum.  That
is, define
\begin{equation}
\bV= \overline{V} \oplus \pi^{*}M
\label{eq:135}
\end{equation}
$\bV$ is a smooth, but reducible, semi--stable, 
 rank $n+m$ vector bundle over $X$ with
structure group $SU(n)\times SU(m)$. Its spectral data can be computed via an
inverse Fourier--Mukai transformation with the result that
\begin{equation}
\bV \longrightarrow (\bC, \bN)
\label{eq:136}
\end{equation}
where
\begin{equation}
\bC =\bar{\cal{C}}\cup m\sigma, \qquad 
\bN =\overline{\cN}\oplus \sigma_{*}M
\label{eq:137}
\end{equation}
Since
\begin{equation}
c_{1}(\overline{V})=c_{1}(\pi^{*}M)=0,
\label{eq:138}
\end{equation}
the Chern classes of $\bV$ are simply the sum
\begin{equation}
c_{i}(\bV)=c_{i}(\overline{V}) +c_{i}(\pi^{*}M)
\label{eq:139}
\end{equation}
for $i=1,2,3$. It follows that 
\begin{equation}
c_{1}(\bV)=0
\label{eq:140}
\end{equation}
\begin{equation}
c_{2}(\bV)=c_{2}(\overline{V}) +kF, \qquad k\in {\mathbb Z}
\label{eq:141}
\end{equation}
\begin{equation}
c_{3}(\bV)=c_{3}(\overline{V})
\label{eq:142}
\end{equation}
This concludes our discussion of reducible $SU(n)\times SU(m)$ vector
bundles. We now turn to the study of small instanton phase transitions
involving 
the pure fiber component of the five--brane class.

Let us begin on the observable boundary brane 
with a stable $SU(n)$ vector bundle $V$ specified by the
spectral cover
\begin{equation}
\cC = n\sigma + \pi^{*}\eta
\label{eq:143}
\end{equation}
and the line bundle $\cN$ over ${\cal{C}}$ whose first Chern class
satisfies~\eqref{eq:27}. It follows from the above discussion that the bulk
space five--brane class $W=W_{B}\sigma+a_{f}F$ does not vanish. In addition,
we will demand that $V$ and $TX$ be such that the fiber component
\begin{equation}
a_{f}F \neq 0
\label{eq:144}
\end{equation}
We will will make no assumptions about the base component, $W_{B}$,
which can be either zero or non-zero, but does not concern us in this section. 

Let us now move the bulk five--brane to the observable boundary brane and
attempt to ``absorb'' the $kF$ part of the five--brane class into the
vector bundle. A full gauge--changing phase transition will occur if we choose
$k=a_{f}$. However, a ``partial'' transition can occur for $k< a_{f}$. 
Our discussion will be similar to that of the previous section,
with the important difference that we must first consider vector bundles over
the base $B$ before lifting them to $X$. With this in mind, define a
rank $m$ vector bundle , $U$, over $B$ by
\begin{equation}
U= \cO_{B} \oplus \dots \oplus \cO_{B}
\label{eq:145}
\end{equation}
with $m$ factors of the trivial bundle $\cO_{B}$ over the base. Henceforth,
for simplicity, we will consider the generic region of moduli space where the
class $kF$ is represented by $k$ separated fibers. Projected onto the
base, this corresponds to $k$ distinct points, $z_{i}$, with
$i=1,\ldots,k$. As in the construction of $\widetilde{V}$ from $V$
and $V_{z}$ in Section~\ref{sec-chirality}, 
we should next specify a line bundle over these points. This
is accomplished by choosing, at each point $z_{i}$, a one dimensional vector
space $U_{z_{i}}$.
The space
\begin{equation}
U_{z}=U_{z_{1}}\cup \ldots \cup U_{z_{k}}
\label{eq:147}
\end{equation}
then defines a line bundle over the base
$B_{z}=\{ z_{1},\ldots,z_{k} \}$ of points. 
Given these two separate vector
bundles, we now attempt to ``weave'' them together by relating them where they
overlap, namely, on $B_{z}$. This is
done  by specifying a surjection
\begin{equation}
\xi:U|_{z_{i}} \rightarrow U_{z_{i}}
\label{eq:148}
\end{equation}
That such a surjection exists will become clear shortly. Given
this relation, one can define a ``bundle'' $\widetilde{U}$ via the
exact sequence
\begin{equation}
0 \rightarrow \widetilde{U} \rightarrow U \rightarrow U_{z}
\rightarrow 0
\label{eq:149}
\end{equation}
It can be shown that because
\begin{equation}
\operatorname{codimension}   B_{z}= 2 > 1,
\label{eq:150}
\end{equation}
then $\widetilde{U}$ is a singular torsion free sheaf, but is not a smooth 
vector bundle. It is easy to show that the general smooth bundle
obtained as a deformation of $\widetilde{U}$ will be stable if and only if
\begin{equation}
2 \leq m
\label{eq:150A}
\end{equation}
Henceforth, we will assume this condition is satisfied.
It is straightforward to compute the Chern classes of
$\widetilde{U}$. They are given by
\begin{equation}
c_{1}(\widetilde{U})=0
\label{eq:151}
\end{equation}
\begin{equation}
c_{2}(\widetilde{U})=k
\label{eq:152}
\end{equation}
\begin{equation}
c_{i}(\widetilde{U})=0, \qquad i\geq3
\label{eq:153}
\end{equation}

We now want to construct the spectral data for the above vector bundles and sheaf
via the inverse Fourier--Mukai transformation. We emphasize that these
transformations are to be carried out in the base space $B$. For simplicity,
we will assume that $B=dP_{9}$. The space $dP_{9}$ is an elliptic 
fibration over $\cp{1} \rm$ with $\pi_{B}:dP_{9} \rightarrow \cp{1} \rm$
and admits a zero section $\sigma_{B}$. Our conclusions, 
however, can be shown to hold generically for any other 
allowed base $B$ \cite{mathpaper}. 
Performing the Fourier--Mukai transformations, we find that
\begin{equation}
U \longrightarrow (m\sigma_{B}, {\mathcal O}_{\cp{1}}(-1)\oplus 
\ldots \oplus {\mathcal O}_{\cp{1}}(-1))
\label{eq:154}
\end{equation}
with $m$ factors of ${\mathcal O}_{\cp{1}}(-1)$ and
\begin{equation}
U_{z} \longrightarrow \bigoplus_{i=1}^{k}
(f_{i}, \cO_{f_{i}}(z_{i}-p_{i}))
\label{eq:155}
\end{equation}
where $f_{i}=\pi_{B}^{-1}(\pi_{B}(z_{i}))$ is the elliptic fiber containing
the point $z_{i}$ and $p_{i} = \sigma_{B}(\pi_{B}(z_{i}))$ is the
origin of $f_{i}$.
In addition, we have
\begin{equation}
\widetilde{U} \longrightarrow (m\sigma_{B} + kf, \widetilde{\cN}_{B})
\label{eq:156}
\end{equation}
where $f \in H_{2}(B,{\mathbb Z})$ denotes the class of the fiber of
the projection $\pi_{B} : B \to \cp{1}$. 
Now, it follows from the Fourier--Mukai transformations and~\eqref{eq:149}
that $\widetilde{\cN}_{B}$ must lie in the exact sequence
\begin{equation}
0 \rightarrow \bigoplus_{i=1}^{k}\cO_{f_{i}}(z_{i}-p_{i})
\rightarrow \widetilde{\cN}_{B} \rightarrow 
{\mathcal O}_{\cp{1}}(-1)\oplus 
\ldots \oplus {\mathcal O}_{\cp{1}}(-1)
\rightarrow 0
\label{eq:157}
\end{equation}
We see that $\widetilde{\cN}_{B}$ must satisfy the condition that
\begin{equation}
\left(\widetilde{\cN}_{B}\otimes
\cO_{B}(-m\sigma_{B})\right)|_{f_{i}}=  \cO_{f_{i}}(z_{i}-p_{i})
\label{eq:159}
\end{equation}
Calculating the first Chern classe of this expression, we find 
\begin{equation} 
c_{1}(\widetilde{\cN}_{B})|_{f_{i}}=m
\label{eq:161}
\end{equation}
for each $i=1, \ldots, k$.
 
We can now examine whether the singular torsion free sheaf,  
$\widetilde{U}$, can
be smoothed out to a stable $SU(m)$ vector bundle on the base, which we
will denote by $\widehat{U}$. The spectral data of $\widehat{U}$ 
can be obtained via
the inverse Fourier--Mukai transformation
\begin{equation}
\widehat{U} \longrightarrow (\widehat{\cC}_{B}, \widehat{\cN}_{B})
\label{eq:162}
\end{equation}
To ensure the stability of 
the bundle $\widehat{U}$, it suffices to choose $\widehat{\cC}_{B}$ 
to be an
irreducible curve in the homology class
\begin{equation}
\widehat{\cC}_{B}= m\sigma_{B}+ kf
\label{eq:163}
\end{equation}
Such a homology class does not necessarily contain an 
irreducible curve. It will contain such a curve if
and only if
\begin{equation}
1 < m \leq k
\label{eq:164}
\end{equation}
which is an important constraint on the choice of $m$. We, henceforth, assume
this condition is satisfied.
We can also show that the spectral line bundle, 
$\widehat{\cN}_{B}$, on $\widehat{\cC}$ must satisfy
\begin{equation}
c_{1}(\widehat{\cN}_{B})= \frac{m}{2}(2k -1 -m)
\label{eq:165}
\end{equation}
The Chern classes of the stable $SU(m)$ bundle $\widehat{U}$
are easily computed. We find that
\begin{equation}
c_{1}(\widehat{U})=0
\label{eq:166}
\end{equation}
\begin{equation}
c_{2}(\widehat{U})=k
\label{eq:167}
\end{equation}
\begin{equation}
c_{i}(\widehat{U})=0, \qquad i\geq3
\label{eq:168}
\end{equation}
which are identical to the Chern classes of the torsion free sheaf,
$\widetilde{U}$, given in~\eqref{eq:151},~\eqref{eq:152} and~\eqref{eq:153}. It
follows that, as far as these Chern classes are concerned, the torsion free
sheaf, $\widetilde{U}$, can be smoothed out to the stable vector bundle,
$\widehat{U}$, without any further restrictions. However, it
remains to check whether the corresponding spectral line bundles satisfy
\begin{equation}
c_{1}(\widehat{\cN}_{B})=c_{1}(\widetilde{\cN}_{B})
\label{eq:169}
\end{equation}
In contrast with the discussion in Section~\ref{sec-chirality}, this
condition does not impose additional constraints. In fact, one can check 
that, on the reducible spectral curve $\widetilde{\cC}$, the sheaf
$\widetilde{N}_{B}$ can be deformed to a line bundle satisfying, in addition
to~\eqref{eq:161}, the condition
\begin{equation}
c_{1}(\widetilde{\cN}_{B})= \frac{m}{2}(2k -1 -m)
\label{eq:170}
\end{equation}
Hence, it follows from~\eqref{eq:165} and~\eqref{eq:170} that
sheaf, $\widetilde{U}$, can always be smoothed out to the stable 
$SU(m)$ vector bundle $\widehat{U}$.

Having defined these bundles and sheafs on the base $B$, we can now pull them
all back to $X$. In particular, $\pi^{*}\widetilde{U}$ is defined on $X$
and, since
\begin{equation}
\operatorname{codimension}   kF= 2 > 1,
\label{eq:171}
\end{equation}
it follows that $\pi^{*}\tilde{U}$ is a singular, torsion free sheaf.
The Chern classes of $\pi^{*}\tilde{U}$ in $X$ are simply the 
pull--back of the Chern classes of $\tilde{U}$ in the base. They are given by
\begin{equation}
c_{1}(\pi^{*}\tilde{U})=0
\label{eq:172}
\end{equation}
\begin{equation}
c_{2}(\pi^{*}\tilde{U})=kF
\label{eq:173}
\end{equation}
\begin{equation}
c_{3}(\pi^{*}\tilde{U})=0
\label{eq:174}
\end{equation}
In addition, it follows from~\eqref{eq:150A} and~\eqref{eq:154} that for
\begin{equation}
2\leq m \leq k
\label{eq:174A}
\end{equation}
the pull--back $\pi^{*}\widehat{U}$ is a stable $SU(m)$
vector bundle on $X$ with spectral cover
\begin{equation}
\hat{\cC}= m\sigma
\label{eq:175}
\end{equation}
and $\widehat{\cN}=\widehat{U}$, which can be 
thought as a deformation of a line bundle over $m\sigma$. 
The Chern classes of $\pi^{*}\hat{U}$ in $X$ are simply the 
pull--back of the
Chern classes of $\hat{U}$ in the base. They are given by
\begin{equation}
c_{1}(\pi^{*}\hat{U})=0
\label{eq:177}
\end{equation}
\begin{equation}
c_{2}(\pi^{*}\hat{U})=kF
\label{eq:178}
\end{equation}
\begin{equation}
c_{3}(\pi^{*}\hat{U})=0
\label{eq:179}
\end{equation}
Finally, it is not hard to establish that, since $\tilde{U}$ can be smoothed
out to $\hat{U}$ in the base, the pull--back torsion free
sheaf, $\pi^{*}\tilde{U}$, can be smoothed out to vector bundle
$\pi^{*}\hat{U}$ on the Calabi--Yau threefold $X$.

The small instanton transition then proceeds as follows. Move the bulk
five--brane to the observable boundary brane. The pure fiber component,
$kF$ of the five--brane class combines with the original stable
vector bundle $V$ on the boundary brane to form a reducible, singular, torsion
free sheaf 
\begin{equation}
\widetilde{\bV}= V \oplus \pi^{*}\tilde{U}
\label{eq:180}
\end{equation}
This small instanton can then be smoothed out to a reducible, but smooth,
semi--stable $SU(n) \times SU(m)$ vector bundle 
\begin{equation}
\widehat{\bV}= V \oplus \pi^{*}\hat{U}
\label{eq:181}
\end{equation}
It follows from~\eqref{eq:139}
and~\eqref{eq:177},~\eqref{eq:178},~\eqref{eq:179} that the Chern classes of
$\widehat{\bV}$ are given by
\begin{equation}
c_{1}(\widehat{\bV})=0
\label{eq:182}
\end{equation}
\begin{equation}
c_{2}(\widehat{\bV})=c_{2}(V)+kF
\label{eq:183}
\end{equation}
\begin{equation}
c_{3}(\widehat{\bV})=c_{3}(V)
\label{eq:184}
\end{equation}

Note that this phase transition changes the structure group on the boundary
brane from $SU(n)$ to $SU(n) \times SU(m)$
where $2 \leq m \leq k$ and, hence, their commutant subgroups in $E_{8}$ also
change. We conclude that such small instanton phase transitions
generically change the unbroken gauge group on the boundary brane.

\subsection*{Summary:}

In this section we have shown the following.

\begin{itemize}

\item Start with a heterotic M--theory vacuum specified by
a stable $SU(n)$ vector bundle, $V$, on the observable boundary
brane with spectral cover
\begin{equation}
{\cal{C}}= n\sigma+ \pi^{*}\eta
\label{eq:185}
\end{equation}
and line bundle $\cN$ over $\cC$ satisfying~\eqref{eq:27},
as well as a five--brane class in the bulk space 
\begin{equation}
W=W_{B}\sigma +a_{f}F
\label{eq:186}
\end{equation}
where $a_{f}F$ is non-vanishing.

\item Now move the five--brane through the bulk space until it touches 
the observable brane and ``detach'' either
all, or a portion, of the fiber component of the five--brane class.
That is, consider $kF$, where $kF$ is either the entire fiber class, for
$k=a_{f}$, or some effective subclass, for $k<a_{f}$.
Leave the pure base component, $W_{B}$ if any, and the rest of the fiber component, 
if any, of the five--brane class undisturbed. One can then define a singular
torsion free sheaf $\pi^{*}\widetilde{U}$ over $X$ associated with $kF$.

\item The original vector bundle, $V$, now combines with 
$\pi^{*}\widetilde{U}$, to form a reducible, singular, torsion free sheaf
\begin{equation}
\widetilde{\bV}= V \oplus \pi^{*}\tilde{U}
\label{eq:187}
\end{equation}
on the observable brane. 
This singular torsion free sheaf is called a small instanton.

\item The small instanton can now be smoothed out into a reducible,
semi--stable $SU(n) \times SU(m)$ vector bundle 
\begin{equation}
\widehat{\bV}= V \oplus \pi^{*}\hat{U}
\label{eq:188}
\end{equation}
where $m$ can be any integer subject to the constraint $2 \leq m \leq k$.

\item The Chern classes of the original vector bundle $V$ and the final bundle
$\widehat{\bV}$ after the phase transition are related by
\begin{equation}
c_{1}(\widehat{\bV})=c_{1}(V)=0,
\label{eq:189}
\end{equation}
\begin{equation}
c_{2}(\widehat{\bV})=c_{2}(V) +kF
\label{eq:190}
\end{equation}
\begin{equation}
c_{3}(\widehat{\bV})=c_{3}(V)
\label{eq:191}
\end{equation}

\item The structure group of the vector bundle changes during the phase
transition from
\begin{equation}
SU(n) \longrightarrow SU(n) \times SU(m)
\label{eq:192}
\end{equation}
where $2 \leq m \leq k$. It follows that the unbroken gauge group on the
boundary brane, the commutant in $E_{8}$ of the structure group,  
also undergoes a transition.

\end{itemize}

These operations define a gauge--changing small instanton phase transition
from one heterotic M-theory vacuum to another involving either all, or part, of
the fiber component of the five--brane class. The base
component, if any, and the remainder of the pure fiber class, if any, 
have not been involved in this transition. In order to make these 
concepts more transparent, we now
present an example.

\bigskip

\noindent
{\bf Example:}

Consider a vacuum specified by $B={\mathbb F}_{1}$, $G=SU(5)$, the
irreducible spectral curve
\begin{equation}
{\cal{C}}=5\sigma + \pi^{*}\eta, \qquad \eta= 24  \cS +36\cE
\label{eq:193}
\end{equation}
and line bundle $\cN$ with $\lambda=\frac{1}{2}$. We showed in Section 5 that
the associated five--brane class is given by
\begin{equation}
W_{B}=0, \qquad a_{f}=132
\label{eq:194}
\end{equation}
and that the number of generations is
\begin{equation}
N_{gen}=336
\label{eq:195}
\end{equation}
Note that the commutant of $G=SU(5)$ in $E_{8}$ is the unbroken gauge group
\begin{equation}
H=SU(5)
\label{eq:196}
\end{equation}
Since $a_{f}\neq 0$, this vacuum satisfies the criterion to make a
gauge--changing small instanton phase transition. To specify this, we must choose
the portion of the fiber component, $kF$, that we want to ``absorb'' during the
transition. Let us choose the entire fiber class by taking
\begin{equation}
k=132
\label{eq:197}
\end{equation}
In this case, the small instanton transition will be to a new smooth, but
reducible, semi--stable vacuum specified by $B={\mathbb F}_{1}$ and
\begin{equation}
\widehat{G}= SU(5) \times SU(m), \qquad 2 \leq m \leq 132
\label{eq:198}
\end{equation}
The entire five--brane class has been absorbed, so 
\begin{equation}
\widehat{W}_{B}=0, \qquad \hat{a}_{f}=0
\label{eq:199}
\end{equation}
Since the third Chern class does not change during the transition, it follows
that
\begin{equation}
\widehat{N}_{gen}=336
\label{eq:200}
\end{equation}
By construction, $SU(m)$ must commute with the structure group $G=SU(5)$.
This implies that 
\begin{equation}
SU(m) \subseteq H =SU(5)
\label{eq:201}
\end{equation}
and, hence, that $m$ is further restricted to satisfy $2 \leq m \leq 5$. 
Moreover, it follows that the commutant, $\widehat{H}$, of $SU(5) \times SU(m)$ 
in $E_{8}$ is the same as the commutant of $SU(m)$ in $H=SU(5)$. It is helpful to
note that the maximal subgroups of $SU(5)$ containing an $SU(m)$ factor are 
\begin{equation}
SU(5) \supset SU(3) \times SU(2) \times U(1), \quad SU(4) \times U(1).
\label{eq:202}
\end{equation}
Let us first consider $m=2$. Using~\eqref{eq:202}, we see that 
\begin{equation}
\widehat{H}=SU(3)\times U(1)
\label{eq:203}
\end{equation}
That is, the small instanton phase transition has changed the gauge
group on the boundary brane from
\begin{equation}
SU(5) \longrightarrow SU(3)\times U(1)
\label{eq:204}
\end{equation}
Similarly, for $m=3$ the gauge group on the boundary brane
undergoes the transition
\begin{equation}
SU(5) \longrightarrow SU(2)\times U(1)
\label{eq:205}
\end{equation}
whereas for $m=4$
\begin{equation}
SU(5) \longrightarrow U(1) 
\label{eq:206}
\end{equation}
Finally, we see from~\eqref{eq:202} that for $m=5$ the gauge
group on the boundary brane changes as
\begin{equation}
SU(5) \longrightarrow {\bf 1 \rm}
\label{eq:206A}
\end{equation}

\smallskip

\

This example highlights two additional properties of gauge--changing small
instanton phase transitions. The first concerns the evaluation of the final
unbroken gauge group. Consider any gauge--changing phase transition in which
the structure group of the vector bundle changes as
\begin{equation}
SU(n) \longrightarrow SU(n) \times SU(m)
\label{eq:207}
\end{equation}
where $2 \leq m \leq k$ and $k \leq a_{f}$. Let $H$ be the commutant of
$SU(n)$ in $E_{8}$ and $\widehat{H}$ be the commutant of $SU(n) \times SU(m)$
in $E_{8}$.  Clearly, since $SU(m)$ must commute with $SU(n)$ we have
\begin{equation}
SU(m) \subseteq H 
\label{eq:208}
\end{equation}
As we learned in the example, this fact generically puts a much stronger 
restriction on $m$. The exact restriction depends on the choice of the
structure group $SU(n)$. In the above example, it tightened the bound on $m$
from $2 \leq m \leq 132$ to  $2 \leq m \leq 5$. Furthermore, it follows
from~\eqref{eq:208} that $\widehat{H}$ must also be the commutant of $SU(m)$
in $H$. This observation facilitates the evaluation of $\widehat{H}$
considerably. For example, let the original stable vector bundle have 
structure group
\begin{equation}
G=SU(4)
\label{eq:209}
\end{equation}
Since $SU(4) \times SO(10) \subset E_{8}$ is a maximal subgroup, it follows
that
\begin{equation}
H=SO(10)
\label{eq:210}
\end{equation}
Now consider a small instanton transition to a reducible vector bundle with
the structure group
\begin{equation}
\widehat{G}=SU(4) \times SU(m)
\label{eq:211}
\end{equation}
where, in general, $2 \leq m \leq k$ for some $k \leq a_{f}$. We see
from~\eqref{eq:208}, however, that
\begin{equation}
SU(m) \subseteq SO(10)
\label{eq:212}
\end{equation}
It is helpful to note that
\begin{equation}
SO(10) \supset SU(2) \times SO(7), \quad SU(2) \times SU(2) \times SU(4),
\quad SU(5) \times U(1)
\label{eq:213}
\end{equation}
are the maximal subgroups of $SO(10)$ containing an $SU(m)$ factor. It follows
that $m$ is further constrained to satisfy
\begin{equation}
2 \leq m \leq 5
\label{eq:214}
\end{equation}
Let us first consider the $m=2$ case. Using~\eqref{eq:213}, we see that
\begin{equation}
\widehat{H}= SO(7), \quad SU(2) \times SU(4)
\label{eq:215}
\end{equation}
depending upon the embedding of $SU(2)$ in $SO(10)$. That is, the small
instanton phase transition has changed the gauge group from 
\begin{equation}
SO(10) \longrightarrow  SO(7), \quad SU(2) \times SU(4)
\label{eq:216}
\end{equation}
Clearly for the $m=3$ case
\begin{equation}
\widehat{H}= {\bf 1 \rm} \Longrightarrow SO(10) \longrightarrow {\bf 1 \rm}
\label{eq:217}
\end{equation}
Similarly, for $m=4$
\begin{equation}
\widehat{H}= SU(2) \times SU(2) \Longrightarrow SO(10) \longrightarrow 
SU(2) \times SU(2)
\label{eq:218}
\end{equation}
and for the $m=5$ case
\begin{equation}
\widehat{H}= U(1) \Longrightarrow SO(10) \longrightarrow U(1)
\label{eq:219}
\end{equation}
We conclude that the gauge bgroup breaking pattern for a gauge--changing 
small instanton phase transition can be computed and is, generically, very
rich.

This observation leads to the second issue regarding gauge--group changing
phase transitions. That is, given the initial data of the stable vector bundle
and the associated five--brane class, can one predict which region of moduli
space the vacuum is in after the phase transition? In particular, can one predict
the final structure group factor $SU(m)$? The answer, for the moment, must be
no. The reason is that, as the five--brane ``collides'' with the boundary
brane, tensionless string states are momentarily created due to the vanishing
tension of wrapped membranes stretched between the boundary brane and the
wrapped five--brane. These states become massive as the small instanton is
smoothed out and can, hence, be ignored after the phase transition. However,
they make it difficult to follow the moduli space trajectory of the vacuum
during the transition itself. We conclude that, presently, we must be content
with constructing the moduli space and specifying the gauge group breaking
patterns of gauge--changing small instanton phase transitions.


\section{Conclusion:}


In this paper, we have given a detailed description of both the mathematics
and physics of small instanton phase transitions associated with the
``collision'' of a bulk space five--brane with a boundary brane. We expect
such collisions and, hence, small instanton phase transitions 
to be an important part of any
realistic particle physics theory derived from the ``brane world''. We have
presented our results within the context of the fundamental brane world 
that arises from M--theory, namely, 
heterotic M--theory \cite{losw1, losw2, nse}. 
However, our results, with relatively minor modifications, are applicable to
any brane world scenario \cite{RS,dim, KT}. 

Specifically, we have shown that upon collision with a boundary brane, part,
or all, of the five--brane is ``absorbed'' by the boundary brane, depending
upon the initial vector bundle and five--brane data. 
The absorbed five--brane is transmuted,
rather catastrophically, into a singular modification of the initial 
vector bundle on the boundary brane, called a small instanton. 
This is then smoothed out to a modified vector bundle that differs
quantitatively from the vector bundle prior to the collision. That is, the
five--brane collides with the boundary brane and disappears, but at the cost
of modifying the vector bundle on the boundary brane. In this paper, 
we have given a precise description of these small instanton phase transitions.

 First, we have shown that if all, or a part, of
the base component of the five--brane class is absorbed during the collision,
then the vector bundle on the boundary brane is modified in such a way that
its third Chern class changes. This implies that the number of
generations of quarks and leptons is different after the phase transition than
it was before. Specifically, we find that
\begin{equation}
N_{gen}(\widehat{V})=N_{gen}(V) + \frac{1}{2}(2\eta + z- nc_{1}(B))\cdot z
\label{eq:220}
\end{equation}
where $V$ and $\widehat{V}$ are the vector bundles on the boundary brane before
and after the transition respectively. Curve $\eta$ and $n$ are 
initial data for the vector bundle on the boundary brane, curve $z$ 
specifies how much of the five--brane class is absorbed during the transition 
and $c_{1}(B)$ partially defines the Calabi--Yau vacuum. The point is that
one can specify this data mathematically and explicitly 
compute the difference in the
number of generations. We showed that the structure group of the
vector bundle does not change during such phase transitions. For these
reasons, we call transitions that involve only the base component of the
five--brane class ``chirality--changing small instanton 
phase transitions''.
At least for the types of vector bundles discussed in this paper, we find that
only for specific initial vector bundle data, namely $\lambda=\frac{1}{2}$,
can chirality--changing transitions proceed. In all other cases 
they are topologically obstructed. 

Second, we have shown that if all, or part, of the fiber component of the
five--brane class is absorbed during the collision, then the vector bundle on
the boundary brane is modified in such a way that the structure group changes.
This implies that the unbroken gauge group on the boundary brane, the
commutant of the structure group in $E_{8}$, is changed by the phase
transition. Specifically, we find that
\begin{equation}
SU(n) \longrightarrow SU(n) \times SU(m)
\label{eq:221}
\end{equation}
where $SU(n)$ and $SU(n) \times SU(m)$ are the structure groups before and
after the phase transition respectively. The values of $m$ are not fixed, but
are constrained to lie in a relatively small interval that can be computed
explicitly. Thus, we can quantitatively compute the unbroken gauge group
structure following the small instanton transition. We showed that the third
Chern class of the vector bundle does not change during such transitions. For
these reasons, we call transitions involving only the fiber component of the
five--brane class ``gauge--changing small instanton phase transitions''.
Unlike the case of chirality--changing transitions, we find that
gauge--changing phase transitions are never topologically obstructed and can
always occur.

In general, small instanton phase transitions involving both the base and
fiber components of the five--brane class can occur. These will then, of
course, involve phase transitions in both the number of chiral families and
the unbroken gauge group. There would appear to be an enormous amount of new,
non--perturbative particle physics and cosmology associated with the small
instanton phase transitions discussed in this paper. We will discuss these
topics elsewhere. 

In this paper, we did not discuss duality between the vacua of heterotic
M-theory and other theories, such as F-theory. We want to point out, however,
that there is an interesting dual 
process in F-theory which follows from the results in our paper. The dual of 
heterotic M-theory on an Calabi-Yau 
elliptically fibered threefold with bundle data is F-theory
compactified on an elliptic Calabi-Yau fourfold.  
We considered chirality changing phase transitions in Section 5. 
In \cite{AnCu}, it is shown that the third Chern class of the vector bundle 
is related to the four-form flux on the F-theory side. 
In type IIB language, 
this flux is the nontrivial NS-NS and R-R three-form field 
background \cite{GVW, DRS}. Since we change the third Chern class of the vector
bundle in chirality changing small instanton phase transitions, 
the dual process in F-theory changes this flux. On the 
heterotic M-theory side, the 
number of five-branes wrapping on a fiber does not change. The dual of the 
5-branes wrapping on a fiber are the three-branes in F-theory \cite{SVW}. 
Hence, the number of three-branes does not change 
in the dual process in F-theory. The anomaly equation in F-theory tells us 
that the sum of the number of three-branes and the flux is proportional to 
the Euler character of the fourfold. The dual process of the small instanton 
transition involving five-branes wrapping on a base curve changes the topology 
of the dual elliptic fourfold in F-theory, since it involves blowdown and 
blowup processes \cite{MoVa, DiRa, Wi3}. Hence, the Euler character of the manifold 
does change before and after the dual process. Thus, we can conclude that 
the F-theory dual process mediates between 
different manifolds with different flux. 
This is difficult to check directly in F-theory, since vacua with 
flux are not well understood. In a similar fashion, 
by further exploration of heterotic M-theory on an elliptic 
Calabi-Yau threefold, 
we can indirectly learn many interesting facts about dual vacua, which are 
worthy of further investigation.

\subsection*{Acknowledgments}

We would like to thank E. Diaconescu, R. Donagi, A. Grassi, K. Intriligator, 
G. Rajesh and E. Witten for useful conversation and discussion. 
B.~A.~Ovrut is supported in part by a Senior Alexander von Humboldt
Award, by the DOE under contract No. DE-AC02-76-ER-03071 and by a
University of Pennsylvania Research Foundation Grant.
T.~Pantev is supported in part by an NSF grant DMS-9800790 and by an
Alfred P. Sloan Research Fellowship.
J. Park is supported in part by U.S. Department of Energy Grant
No. DE-FG02-90-ER40542.

\end{document}